\documentclass[iicol,sn-basic]{sn-jnl}

\usepackage{graphicx}%
\usepackage{multirow}%
\usepackage{amsmath,amssymb,amsfonts}%
\usepackage{amsthm}%
\usepackage{mathrsfs}%
\usepackage[title]{appendix}%
\usepackage{xcolor}%
\usepackage{textcomp}%
\usepackage{manyfoot}%
\usepackage{booktabs}%
\usepackage{algorithm}%
\usepackage{algorithmicx}%
\usepackage{algpseudocode}%
\usepackage{listings}%
\usepackage[colorinlistoftodos]{todonotes}
\usepackage{svg}
\usepackage{graphicx}
\usepackage{subcaption}
\usepackage{pdfpages}
\theoremstyle{thmstyleone}%
\theoremstyle{thmstyletwo}%

\theoremstyle{thmstylethree}%
\usepackage{url}
\newfloat{imagetable}{h}{}
\floatname{imagetable}{Table}
\restylefloat{imagetable}

\begin{document}

\title[Article Title]{Adaptive Framework for Ambient Intelligence in Rehabilitation Assistance}

\author{\fnm{G.} \sur{Baranyi$^1$}}%\email{baranyi@inf.elte.hu}

\author{\fnm{Zs.} \sur{Csibi$^1$}}%\email{iiauthor@gmail.com}

\author{\fnm{K.} \sur{Fenech$^1$}}%\email{iiiauthor@gmail.com}
\author{\fnm{Á.} \sur{Fóthi$^1$}}%\email{iiiauthor@gmail.com}
\author{\fnm{Zs.} \sur{Gaál$^2$}}%\email{iiiauthor@gmail.com}
\author{\fnm{J.} \sur{Skaf$^1$}}%\email{iiiauthor@gmail.com}
\author{\fnm{A.} \sur{Lőrincz$^1$}}%\email{lorincz@inf.elte.hu}

\affil{$^{1}$ \orgdiv{Dept. of Artificial Intelligence}, \orgname{Eötvös Loránd University}, %\orgaddress{\street{Street}, \\
\city{Budapest}, \postcode{H-1117}, %\state{State}, 
\country{Hungary}}

\affil{$^{2}$ \orgname{Emineo Private Hospital}, %\orgaddress{\street{Street}, \\
\city{Budapest}, \postcode{H-1016}, \country{Hungary}} %\state{State}, \country{Hungary}}


\abstract{This paper introduces the Ambient Intelligence Rehabilitation Support (AIRS) framework, an advanced artificial intelligence-based solution tailored for home rehabilitation environments. AIRS integrates cutting-edge technologies, including Real-Time 3D Reconstruction (RT-3DR), intelligent navigation, and large Vision-Language Models (VLMs), to create a comprehensive system for machine-guided physical rehabilitation. The general AIRS framework is demonstrated in rehabilitation scenarios following total knee replacement (TKR), utilizing a database of 263 video recordings for evaluation. A smartphone is employed within AIRS to perform RT-3DR of living spaces and has a body-matched avatar to provide visual feedback about the excercise. This avatar is necessary in (a) optimizing exercise configurations, including camera placement, patient positioning, and initial poses, and (b) addressing privacy concerns and promoting compliance with the AI Act.
The system guides users through the recording process to ensure the collection of properly recorded videos. AIRS employs two feedback mechanisms: (i) visual 3D feedback, enabling direct comparisons between prerecorded clinical exercises and patient home recordings and (ii) VLM-generated feedback, providing detailed explanations and corrections for exercise errors.
The framework also supports people with visual and hearing impairments. It also features a modular design that can be adapted to broader rehabilitation contexts. AIRS software components are available for further use and customization.}




\maketitle

\section{Introduction}\label{sec1}
Technology advances rapidly and is influencing our daily lives. Smart homes and devices are increasingly available and offer various forms of support.  In situations such as physical rehabilitation, there is a significant need for remote assistance, such as tele-rehabilitation, especially when distance from experts and mobility challenges can pose serious barriers to traditional on-site rehabilitation.

The term Ambient Intelligence (AmI) emerged around 1990, inspired by the European Union's 5th Framework Programme on "Information Society Technologies" (IST). AmI integrates artificial intelligence, ubiquitous computing, and human-centered design to create adaptive environments using technologies such as sensor networks and context-sensitive systems. Ambient Assistive Living (AAL) systems, such as those discussed by \cite{cicirelli2021ambient}, support elderly, people with chronic illnesses, disabilities, and cognitive impairments.

\begin{figure*}[t]
    \centering
    \includegraphics[width=1.0\textwidth]{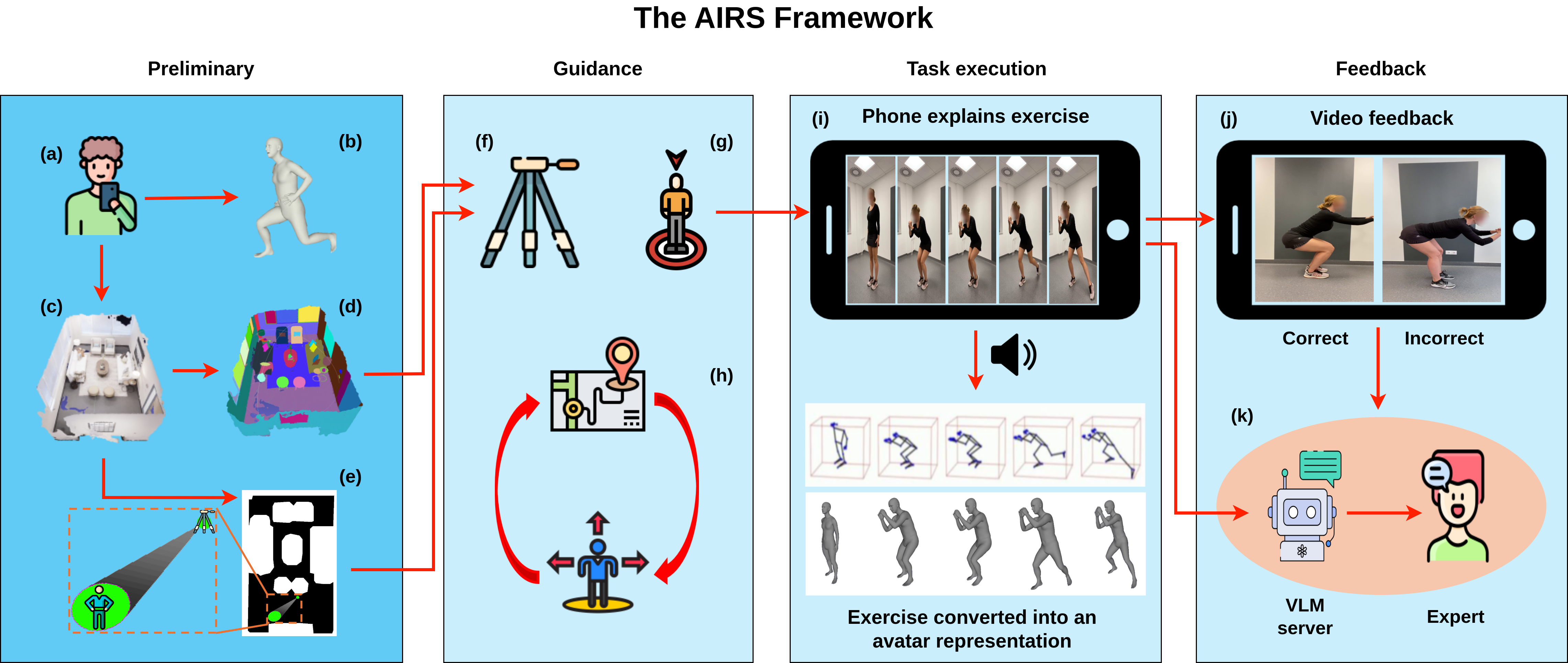}
    \caption{(a) Patient using a smartphone, (b) pseudonymized 3D reconstruction of the patient using body mass index based estimation, (c) reconstruction of the patient's home environment using 3D Gaussian Splat models, (d) coloured semantic map for route planning to the practice space, (e) computation result of the optimal position of the tripod and  of the patient, (f) computation result of the optimal pose of the camera and (g) the patient, (h) navigation module guiding the patient in setting up the tripod, adjusting the camera, setting the initial state, (i) visual and verbal instructions via mobile phone to start and perform the exercise, start video recording, and numerical procedures to eliminate the background, estimate the patient's skeleton during the exercise and convert it into a meaningful avatar representation using a body mass index-based estimation to protect personal information, (j) presenting synchronized versions of correct videos of exercises recorded in the clinic and those recorded at home, respectively for visual feedback, and (k) a large vision-language model that provides verbal feedback on exercise performance and on perceived and suspected errors  to the therapist.     The design has resources from Flaticon.com. 
}
    \label{fig:framework}
\end{figure*}

Our focus is on aiding physical rehabilitation patients, addressing environmental challenges in smart homes. We want to extend the advantages of telerehabilitation by overcoming its limitations concerning human assistance. Advances in 3D reconstruction, large Vision Language Models (VLMs), and traditional AI methods, such as rule-based systems, speech-to-text algorithms, and smartphones enable spatial awareness and intuitive interaction. Using these technologies, we developed and evaluated the Ambient Intelligence Rehabilitation Support (AIRS) framework (see Fig.~\ref{fig:framework}).

We study post-operative rehabilitation for Total Knee Replacement and include the most important components, such as methods, algorithms, and software that we detail in the Supplementary Materials. AIRS is general; it is not restricted to TKR. In addition, it can provide feedback to users with visual or hearing impairments.

Our contributions contain the following components:
\begin{itemize}
    \item We present the AIRS framework, integrating AI technologies such as real-time 3D reconstruction and Vision Language Models (VLMs) for machine-guided physical rehabilitation, including modeling of the 3D environment, exercise optimization, and visual feedback.
    \item Temporally adjusted visual feedback highlights posture differences between clinical and home recordings for the therapist and patient, resepctively.
    \item We assess pre-trained Vision-Language Models (VLMs) for generating verbal feedback on error detection and corrections on observed motion. Although VLMs may be efficient in finding errors in home recordings, they may fall short in critical cases. On the other hand, they can save time for the therapist.
    \item We replace the patient with a matching avatar for privacy reasons. This solution allows data collection, enables AI systems to analyze body motion and rehabilitation progress, optimize exercises, and design future rehabilitation sessions.   
    \item For homecare, we offer modular cross-platform AIRS software compatible with iOS and Android, adaptable for various applications.
    \item We underscore flexibility of AIRS in guiding blind or hard-of-hearing individuals in rehabilitation exercises with minimal code adjustments.
\end{itemize}

The remainder of the paper is structured as follows. In Sect.~\ref{sec2}, we discuss ambient intelligence solutions for physical rehabilitation and healthcare, together with smartphone-based rehabilitation approaches that are closely related to our work. Next, in Sect.~\ref{sec3}, we present the system pipeline in detail, beginning with an introduction to the data set that we collected and used in this study. Section~\ref{sec5} presents results on the precision of errors detected by VLMs and studies on a subset of the original dataset how much vision language models can help physical therapist. It also addresses potential future improvements.
Finally, in Sect.~\ref{sec6}, we conclude with a brief summary of our work. The project page can be found at \href{https://baranyigabor.github.io/ambient-rehab/}{this https URL}. The code will be made available upon publication.

\section{Motivation and related works} \label{sec2}

The home rehabilitation market is growing rapidly. The growth is driven by the aging population and their preference for home care. Conditions such as arthritis, stroke, and cardiovascular disease, along with the increasing geriatric needs, increase the demand for rehabilitation services.

However, a global shortage of skilled physical therapists and the increasing need of home care pose challenges, leading to suboptimal quality of care. Innovations like telehealth are facilitating home-based therapy, but meeting the growing demand will require further advancements in smart, ambient systems and corresponding workforce development.

Visually impaired or hard-of-hearing individuals pose additional challenges to AmI in such applications, e.g., in guiding their therapy sessions or navigating and recording exercises in their environments. We argue that real-time 3D reconstruction and VLMs could offer a comprehensive support system for home rehabilitation for everyone, including people with sensory impairments.

In our work, we present an intelligent application for home rehabilitation that can promote independence and improve the quality of individual lives with sensory impairments. We will also discuss this specific aspect.

\subsection{Ambient intelligence solutions for rehabilitation}

Multiple studies have investigated the integration of ambient intelligence (AmI) and machine learning in healthcare and physical rehabilitation. We list a few more recent samples from this vast literature. The interested reader is referred to the work of \cite{patel2025hospital} for further information on personalized, efficient, and accessible AmI and the related problems. \cite{javed2020collaborative} introduced the "Collaborative Shared Healthcare Plan", which monitors daily activities and fosters collaboration among healthcare providers and caregivers of patients. \cite{kim2020neural} proposed a neural network-based model that improves health predictions by adapting to changing user conditions. \cite{passias2024biologically} propose an energy-efficient movement recognition system for Ambient Assisted Living using Spiking Neural Networks (SNNs), achieving high accuracy while reducing energy consumption compared to deep neural networks (DNNs).

In the context of cognitive rehabilitation, the work of \cite{oliver2018ambient} emphasizes the use of interactive systems equipped with smart sensors for remote monitoring. In their article, \cite{han2023ai} presents an AI system that combines virtual reality and exoskeletons for the rehabilitation of the elderly people. \cite{roda2015multi} describe a multi-agent architecture that promotes personalized therapies for older adults with motor impairments.

\cite{s21196692} presents a wearable system for fitness and rehabilitation, using sensors such as accelerometers and gyroscopes to collect motion data and provide real-time feedback for posture correction. Its portable design, user-friendly interface, and progress tracking enhance exercise effectiveness, prevent injuries, and personalize routines, representing advances in wearable fitness technology. The cost and complexity of the wearable system can form severe barriers for the elderly, and visual feedback is not available in this type of solution.  

In line with our work, \cite{zhao2024smartphone} developed and carefully tested a smartphone application for TKR. \cite{kryeem2025action} developed AI tools for the evaluation of exercises. In our work, we also target additional tasks helping home-based rehabilitation, such as navigation strategies in unknown environments, the optimal arrangement plan, including aspects of high-quality recordings such as camera placement, the patient's initial pose or video-based comparisons, and the evaluation of the effectiveness of VLM in providing feedback to both the expert and the patient.

It follows that the AIRS framework consists of many components. A brief review of related work for each component is provided in the Methods section to enhance readability.

\section{System design and methods}\label{sec3}

Our proposed framework goes beyond aiding exercise execution in rehabilitation. It also navigates the user, assists with camera placement, ensures proper video capture, and can provide notes about errors to the therapist, using VLMs. We test the quality of error correction outputs of VLMs. The results are promising for one of the VLMs, but far from the quality required to directly help the patients. We study and estimate the capabilities of the components.

\subsection{Dataset}
\label{sec:llm:data}

The dataset used in this study consists of 263 videos of 43 different physical therapy exercises.  These data were collected at the Emineo Private Hospital, Budapest, see \citep{baranyi_ai_rehab}. Recording happened with a single camera. Each exercise includes a correct performance by a therapist and several videos showing specific errors, accompanied by descriptions. The number of videos containing errors for the different exercises is between 3 and 10 depending on the complexity.

The dataset includes patient discussions, therapist's queries, videos, exercise descriptions, error types, and corrections. Here, we extend and improve our previous work. We used all components of the database, except patient-therapist discussions. Concerning the LLM-VLM methods, errors and their descriptions were incorporated into the prompts, the therapist provided the description of corrections that served as ground-truth data for us. In addition, images were provided to the LLM-VLM system.

\begin{figure}[t!]
    \centering
    \includegraphics[width=0.485\textwidth]{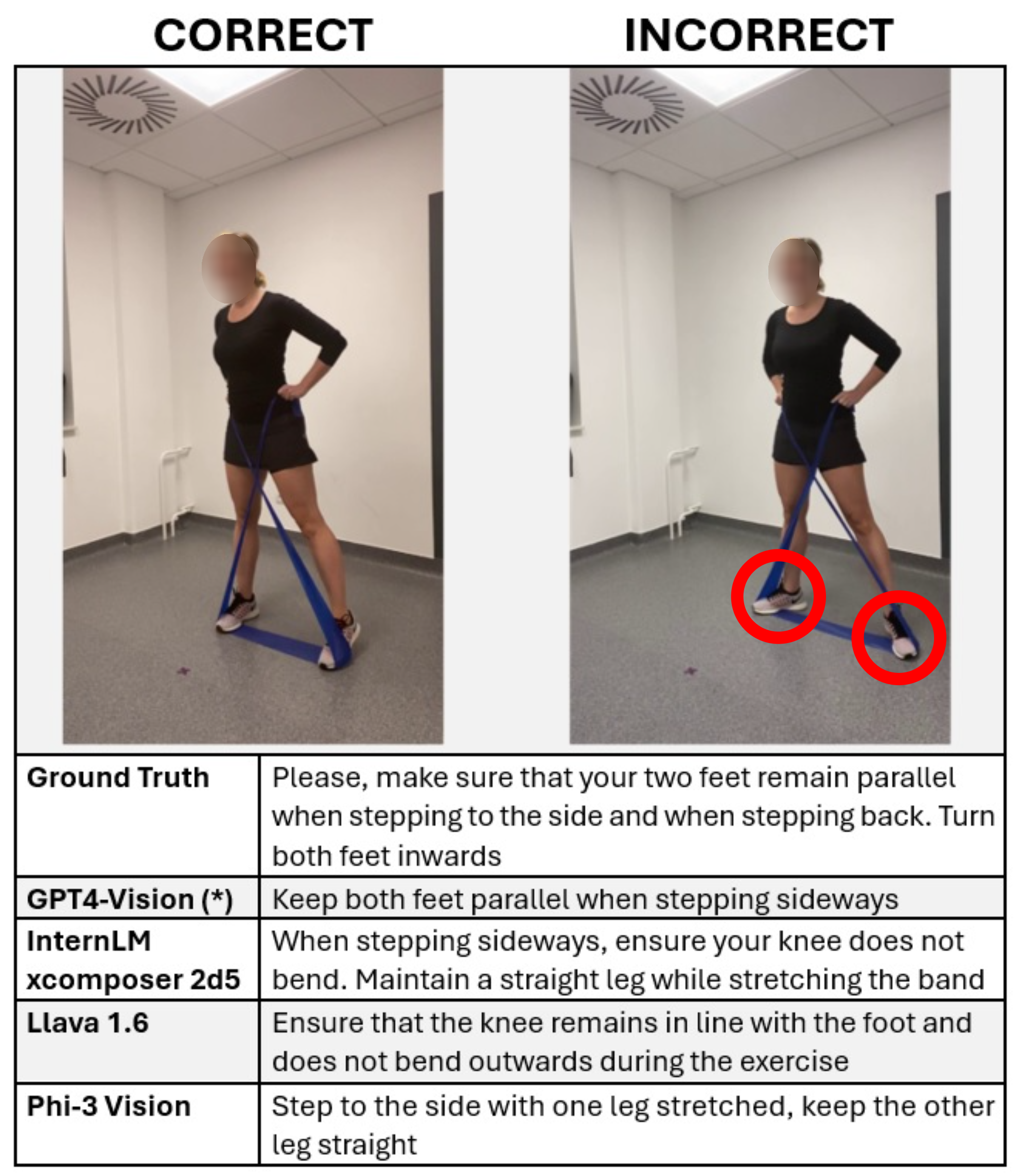}
    \caption{Ground truth and error correction suggestions by four different VLMs. For details, see text.}
    \label{fig:example_corrections}
\end{figure}

An example is shown in \autoref{fig:example_corrections}. The `patient (the physical therapist herself in  our case) performs the exercise correctly in the clinic and makes a typical but relevant mistake at `home'. The ground truth from the expert and the suggestions for corrections from four models are shown. Only GPT4-Vision gives a reasonably good response. We note that a totally correct VLM response is not possible due to estimation errors in the angle of the feet.

\subsection{Avatar creation}

The avatar creation step is indicated by Fig.~\ref{fig:framework}(b) of our framework. This creation of the patient avatar is one step towards satisfying the constraints of the EU Artificial Intelligence Act\footnote{\url{https://eur-lex.europa.eu/legal-content/EN/TXT/?uri=celex:52021PC0206}} and to allow data collection at home. It is also necessary to find the optimal place for the exercises. 

Single-shot models such as Hybr-IK \citep{li2023hybrik}, ExPose \citep{choutas2020monocular}, and NIKI \citep{li2023niki} recover 3D meshes from a single image, but poor camera angle, occlusion, and suboptimal lighting may pose problems. 

Automated 3D model generation techniques like Neural Radiance Fields (NeRF) \citep{mildenhall2021nerf} and Gaussian Splatting (GS) \citep{kerbl20233dgaussiansplattingrealtime} can develop detailed 3D models with textures from video. However, these methods require an additional person to record the video and involve a manual skeleton parenting process to annotate joints and dependencies in Blender\textsuperscript{\textregistered}.

One can use Blender's BMI (Body Mass Index) function \citep{choutas2020monocular} and height data to create avatars, but Blender can be complex for most users.

If the skeleton of the person can be estimated, then the parameterized Skinned Multi-Person Linear (SMPL) model \citep{SMPL2015} can be used. SMPL maps skeletal configurations onto an SMPL structure using theta parameters for poses and beta parameters for personal attributes such as height. We used the skeleton-based model (SKEL) \citep{keller2023skin} that incorporates biomechanical constraints and refines the SMPL model.

Depending on the application, for example, when the patient records the exercise video, we can extract the skeleton data. MeTRAbs \citep{metrabs_upgraded}, a 3D human pose estimation method, utilizes robust metric-scale truncation heat maps and can produce various skeleton representations in two and three dimensions, which we used for our estimations.

\subsection{3D reconstruction of the environment}

This step is indicated in Fig.~\ref{fig:framework}(c) of our framework. 
Reconstruction of the 3D environment is key to modeling the unoccupied area of the environment and optimizing exercise and camera positioning. We distinguish two scenarios. In the offline scenario, the patient or an assistant records the environment by following a few instructions and capturing the relevant parts using a smartphone. In the online scenario, the system builds the 3D model as data is collected. The camera parameters are estimated in near-real time. It is more user-friendly as it provides immediate visual feedback, but requires backing computational power. If this is not the case, then NeRFStudio \citep{nerfstudio} and the pre-recorded video method --~detailed below~-- should be preferred.

NeRFStudio supports both NeRF and Gaussian Splatting (GS), the most important reconstruction algorithms. NeRF works with neural networks for high-fidelity 3D reconstructions. GS developed by \cite{kerbl20233dgaussiansplattingrealtime} 
 combines point clouds with smooth surfaces more suited for real-time applications. 

NeRFStudio also employs COLMAP \citep{schonberger2016structure} to calculate the camera intrinsic and extrinsic parameters through Structure from Motion (SfM), which involves feature extraction, feature matching steps to achieve accurate transformation matrices of each recorded frame for the reconstruction step. The 3D model is the output of COLMAP. Features are used in the online camera pose estimation to be detailed in Sect.~\ref{localize}

In the real-time reconstruction pipeline, images from the camera are streamed  to a server, and the localization tool computes the odometry data. Odometry and images are fed to NeRFStudio through NeRFBridge \citep{yu2023nerfbridge}, which facilitates real-time communication with NeRFStudio using ROS2 communication protocols. Model training starts with a few images, and NeRFBridge gradually adds new images to refine the 3D model. The model improves and can be tracked in real-time. 

\subsection{Path planning and its prerequisites}\label{sec4}

Verbal navigation could use egocentric directions and distances or the semantic map of the environment. The former is much less convenient, and the latter works well if visible but relatively distant positions are mentioned (see our previous study \citep{baranyi_ai_rehab}) as illustrated in Fig.~\ref{fig:framework}(d). The three prerequisites of path planning include the computation of the \emph{optimal final positions} (Fig.~\ref{fig:framework}(e)) and \emph{poses} of the camera (Fig.~\ref{fig:framework}(f)) and the user (Fig.~\ref{fig:framework}(g)) followed by the real-time estimation of the user's actual location and viewing angle, that is, real-time localization. Having these information pieces, path planning and finally navigation can start (Fig.~\ref{fig:framework}(h)).

We start with the optimization task, followed by the solution to the localization problem. The last point is about path planning.

\subsubsection{Identifying target area and poses}

We identify an area suitable for performing the target exercise safely and comfortably. The camera should be placed at an optimal distance and angle relative to the patient (Fig.~\ref{fig:framework}(e)) taking into consideration the viewing angle of the camera, the sizes of the avatar, visibility of the relevant joints to reduce errors in single-image pose estimations, and the space (volume) required by the specific exercise.

We calculate the maximum spatial requirements needed for the safe execution of the exercise.
We use the avatar model matching the user and
combine the series of pose estimations of the
frames of each video taken under the quality control of the physical therapist while the patient
was performing the exercise. 

We project the estimated poses to the floor and calculate the convex hull of the projections. The boundary of the convex hull is then approximated
by the minimal ellipsis that completely covers the hull and thus the entire movement area. We get the necessary volume by adding the third dimension, the maximal height needed for the exercise. 

The optimal distance of the camera depends on its viewing angle. All surface points of the visible part of the volume defined above should be visible from the optimal viewing angle. This way we define a volume that contains the ellipsoid and the camera. This volume should be placed on the floor in such a way that the volume should not intersect with any objects.
    
Once the exercise boundaries and camera constraints are defined, it is important to convert the measurement units (e.g., meters) into pixels, using the same conversion factor applied to the top-down map of the environment (Fig.~\ref{fig:framework}). Two matrices are generated: the first represents the top-down map with occupied and empty spaces, and the second represents the exercise boundaries. To identify a suitable location for the exercise, we employ a sliding window approach, which scans the entire environment matrix. By sliding the exercise boundary matrix across the top-down matrix and applying a logical AND operation, we can detect empty regions. When a match is found, that area is considered suitable for exercise execution. The search continues across the entire matrix to identify all possible solutions. If no viable position is found, the exercise boundaries can be rotated to different angles and re-evaluated. 

For each viable solution, we compute a score function that prioritizes positions with more free space around the exercise boundaries. After ranking the scores, the optimal one can be selected.

\subsubsection{User localization}\label{localize}

Determining user's and object's pose and location in real-time is essential for navigation and arranging the optimal starting poses. To address this challenge, we extended the approach presented in Hierarchical Localization (Hloc) \cite{sarlin2019coarse} by feeding a stream of images and identifying the best match for each new image starting from among the preceding image in the stream. This approach reduces the number of images required to match new queries and decreases the overall processing time to meet real-time requirements. The method works on devices with limited computational power.

This approach is divided into two main phases. The first phase aims to construct an offline 3D reference point cloud of the environment. In the second phase, the points in the reference point cloud are used to determine the location and orientation of the new camera frame. This slight modification and the other details of our method can be found in the Supplementary Materials.m

For feature extraction and matching, an independent deep learning algorithm performs the two steps, namely feature extraction and feature matching. 
Superpoint \citep{detone2018superpoint} is used for feature extraction, while LightGlue \citep{lindenberger2023lightglue} is selected for feature matching. Both are fine-tuned to account for the accuracy and speed trade-off, enabling efficient real-time applications.

The path planning step uses the current position and the target area. The approach is an extension of our previous work \citep{baranyi_ai_rehab}. We used Habitat \citep{szot2021habitat} for testing and replaced the semantic information of Habitat with the EmbodiedSAM method \citep{xu2024embodiedsam}, which utilizes Segment Anything 2 \citep{ravi2024sam2segmentimages} SAM2 in short, for semantic labeling over the given RGB images. Our algorithm is a rule-based system that collects semantic information from EmbodiedSAM, identifies obstacles and possible routes, allowing the generation of navigation instructions at any instant subject to the position and pose of the user. These components, except EmbodiedSAM, are classical methods. Details can be found in the Supplementary Materials.

\subsection{Navigation}\label{sec4_}

After obtaining a 3D model of the environment and the user, the desired path at any instant, the next step is navigating the user within the environment. The \emph{navigation module} guides the user from the current position to a target area through the following two steps:

\begin{itemize}
    \item \textbf{Approximate User localization} We assume that the camera is a smartphone and is held in front of the user with the screen facing the user like and the rear camera is active. Then the localization of the camera described in Sect.~\ref{localize} approximates the position and direction of the user.
    \item \textbf{Giving instructions and path refinements} All instructions are egocentric, provided from the person's perspective. This is a simple computation as we have the direction of the path and the direction of the user at any point. Directional instructions can be displayed on the phone with an option for voice guidance. It is more convenient for most sighted individuals if visible remote objects are told as intermediate locations along the path shown in the Habitat simulated navigation tests \citep{baranyi_ai_rehab}. The patient can request new instructions by pressing a volume button. Then the navigation module recalculates and the instruction to the target area at the actual, e.g., at the starting point. 
\end{itemize}

\subsection{Performing and evaluating the exercises}
When the patient and the camera are properly positioned, the execution of the task can be started together with the recording of the videos.

\subsubsection{Processing the videos}
\label{processing}

When the video recording is finished, the first step is to extract the skeleton data. Here, we used two different approaches. In the first approach, standard SMPL skeleton coordinates are extracted using MeTRAbs. The second approach consists of two steps: first, the SMPL parameters were extracted for every frame of the video using HybrIK-X, then the SMPL parameters were refined with SKEL to obtain the biomechanically accurate skeleton parameters for every frame. 

The recorded videos are then synchronized with the corresponding reference videos, which are played and recorded under the supervision of the therapist as part of normal clinical practice. Note that the player is the same on both videos, so no additional matching is necessary. Dynamic Time Warping (DTW) was sufficient in aligning our relatively low-dimensional time series data, which could be either calculated for joint angles (e.g. left ankle - center of hip - right ankle) or SKEL pose parameters (rotation of each joint).

From the synchronized skeletons and image pairs, the one containing the most significant error is selected based on the Mean Square Error or Mean Absolute Error using either SKEL pose parameters or joint angles. In the VLM studies, these selected image and skeleton pairs are then used in various prompts with different methods; see below.

\subsubsection{Prompt Building}

Large vision language models are advancing in diverse tasks like visual question answering and reasoning but require large databases, long and memory-consuming trainings. 

To address domain-specific challenges, such as physical rehabilitation --~where data are sparse and the VLM must compare specific posture parameters in two recordings, highlight minor differences, and provide corrective instructions~--prompting techniques can be highly effective.

We used various input types with an additional feature in the zero-shot Chain-of-Thought (CoT) prompting technique suggested by \cite{wei2022chain}. Our goal was to see to what extent VLM could help the physical therapist. As we show, the answer presently is yes, VLMs are useful for helping the therapist by pointing to certain errors. VLM's suggestions for correction instructions can be of high-quality but this is not the case always that we discuss later. 

In our experiments, we compared the effectiveness of different prompts, such as the image pairs selected by DTW, SMPL-based body structure prompts, Error List (EL) prompts for contextual help, and text-based Body Region (BL) prompts for spatial help. Prompts can have 250 lines. For details, such as SMPL skeleton coordinates, domain-specific knowledge, e.g., exercise descriptions, the list of errors, and the list of specific body regions, see the Supplementary Materials.

\subsubsection{VLM evaluations}

To evaluate the responses of the VLM model, two approaches were used to compare the generated correcting instructions with the ground truth ones: embedding-based similarity and semantic similarity accuracy.

Here, semantic similarity \emph{accuracy} means that, according to a VLM model, the meaning of the ground truth and the generated sentence match each other, a binary classification task. VLMs were prompted to assess the semantic similarity accuracy. The results are given in percentage for the entire database and for different input types.

We can assess the internal representation for certain VLMs. These embeddings can be compared for ground truth and suggested corrections.  We used cosine similarity and averaged the similarities for each method. For input sentence embedding, we used the representation of the last hidden state of the software LlaMa-3-8B-Instruct. It is a 4096-dimensional vector.

%---------------------------------
\section{Results}\label{sec5}
%---------------------------------

Our work provides a methodology, algorithm suite, and software modules that prioritize privacy, ensure AI Act compliance, enable remote collaboration between patients and caregivers, and improve on traditional telerehabilitation. The RT-3DR method enables real-time navigation by constructing and refining an environment model, localizing the user, planning an optimal path, and offering step-by-step guidance. Further details are available in the Supplementary Materials.

Here, we focus on two key aspects for later discussions: (a) whether the precision of single-camera body pose estimation is sufficient for providing visual feedback to the patient, and (b) whether current VLMs are effective enough to provide verbal feedback on error corrections to the patient or to reduce the physical therapist's workload.

We evaluated 43 videos about different TKR rehabilitation exercises and 263 videos that showed potential errors, all of which were executed by the expert physical therapist.   

%---------------------------------
\subsection{Smartphone frontend}
%---------------------------------

\begin{figure}[t!]
    \centering
    \includegraphics[width=0.4\textwidth]{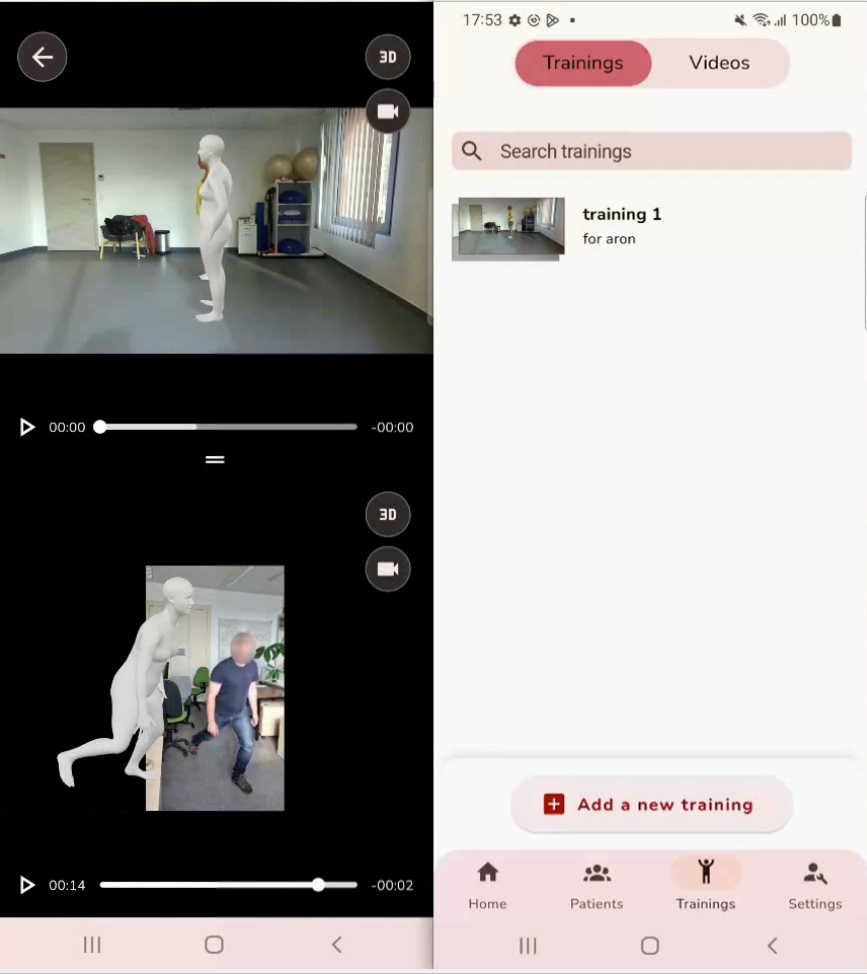}
    \caption{Two smartphone frames. Left: a sample exercise at the clinic and an excercise at home showing options: video and/or avatar. Right: state of the ongoing exercise.}
    \label{fig:smartphone}
\end{figure}

The code is in flutter, a multi-platform open-source UI software development kit from Google. It can be adapted for various applications. Two screenshots are shown in Fig.~\ref{fig:smartphone}

%---------------------------------
\subsection{Precision of single camera recordings for error detection}
%---------------------------------

The DTW method was used to synchronize the error videos, followed by selecting the most critical error frames for each video using the mean square error or the mean absolute error in 263 error videos spanning 43 different exercises. Two individuals reviewed the selected frames to verify whether the error was present relative to the ground truth video. They agreed in 245 cases and reached a consensus in 18 others. Of the 263 videos, DTW selected frames in which they could see the error but did not see the error in 29 cases.

\begin{figure}[t]
    \centering
    \includegraphics[width=0.9\linewidth]{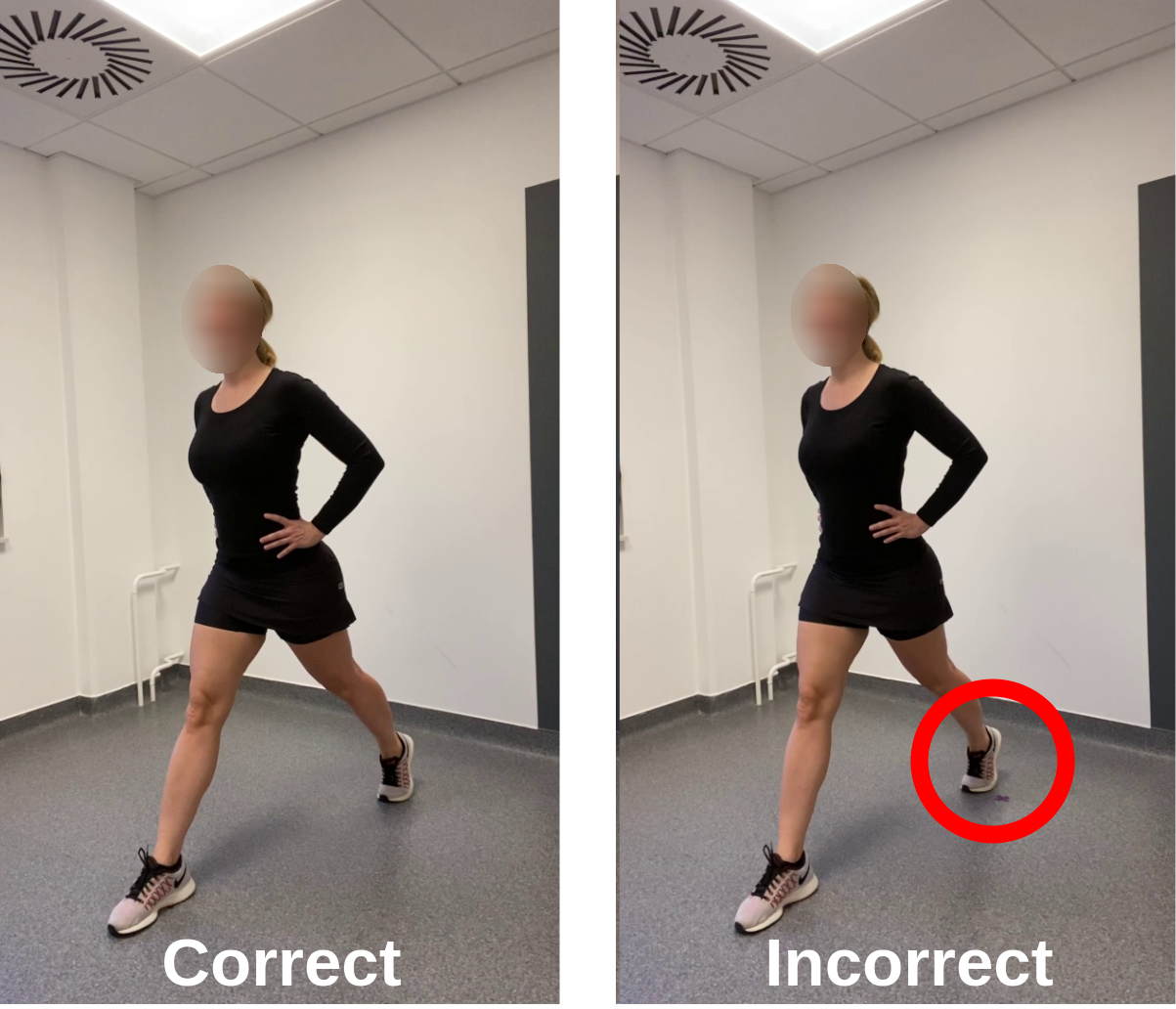}
    \caption{\textbf{Hard to detect error:} The back foot is overly aligned with the front foot, reducing stability and making it harder to maintain balance.}
    \label{fig:hard_error}
\end{figure}

\begin{figure}
    \centering
    \includegraphics[width=0.9\linewidth]{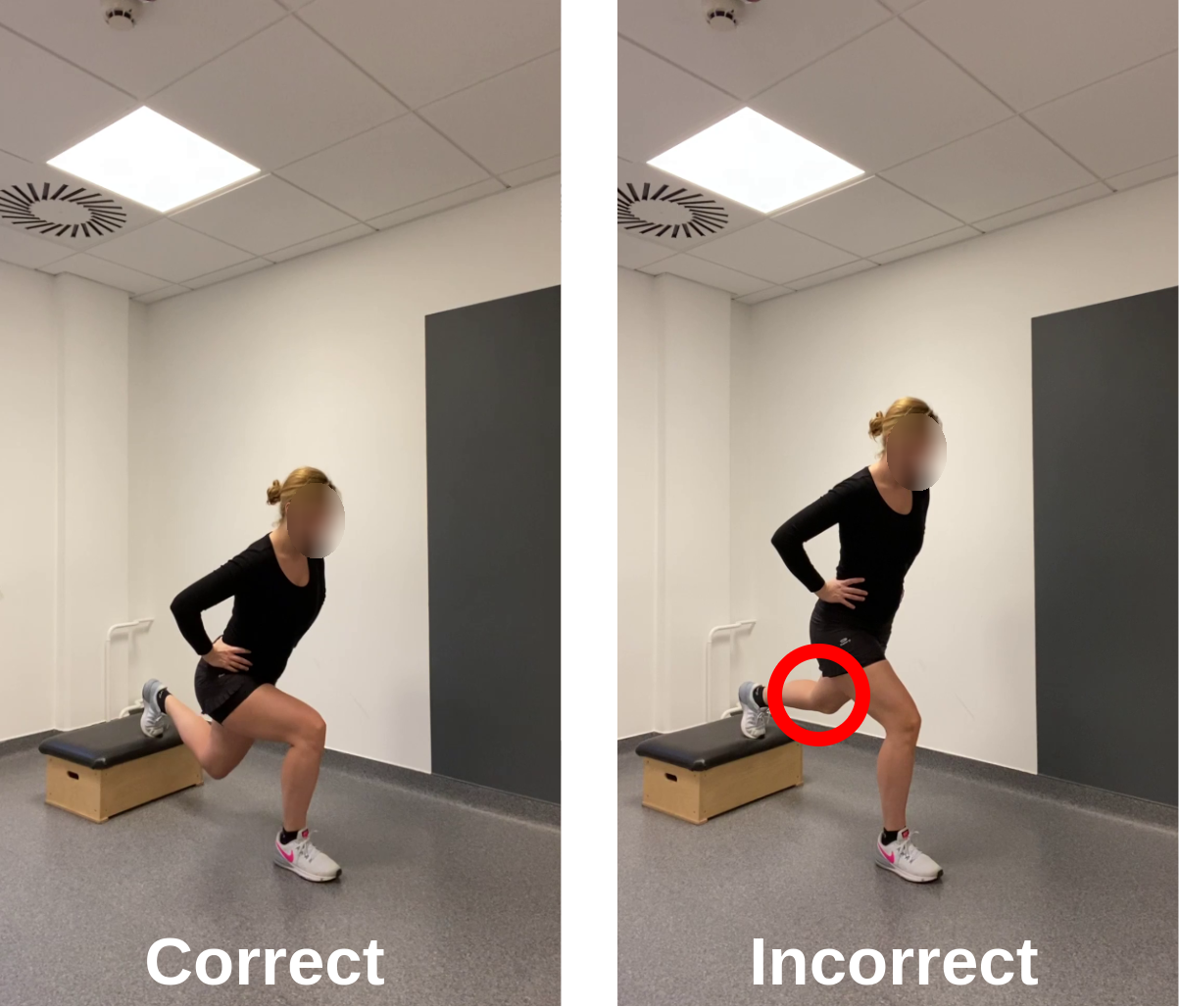}
    \caption{\textbf{Easy to identify error:}  Does not bend the knees deeply enough.}
    \label{fig:easy_error}
\end{figure}

Some errors among the 29 cases are almost undetectable. One case is shown in Fig.~\ref{fig:hard_error}. The right image of these DTW-synchronized frames shows a challenging balance pose with a subtle difference (a spot) near the left foot. Such small errors require precise RGBD recording for detection, as losing balance can be dangerous and strain the body. However, the ratio of detected errors is as high as 89\% and can be highlighted in the recorded videos as in Fig.~\ref{fig:easy_error}. 

\subsection{Testing Vision Language Models}

We selected 15 different error videos that had the highest number of potential errors. We run several studies (see the Supplementary Materials). Two of them are shown here. We used GPT-4 Vision to compare the instructions generated by the different VLMs. We used the instruction received from GPT-4 Vision as input together with the ground truth of the expert and asked if their meanings were the same. Table~\ref{tab:accuracies} shows the results. In one of the cases, we provided the "Error List", i.e. the description of the different errors as prompts, and the task was to choose the right one. In another case, the relevant body region was noted in the prompt. The other two cases had neither of these information pieces and both of them, respectively. The Table shows the results for image, SMPL, and image + SMPL prompts, too.  The SMPL and Body Region prompts together are better by one more hit (out of 15) than when we add the Error List prompts. This peculiar exception is against the general impression of the Table that adding prompts increases the number of hits up to 9 (60\%). 
\begin{table}[b!]
    \centering
    \begin{tabular}{c}
         \includegraphics[width=0.94\linewidth]{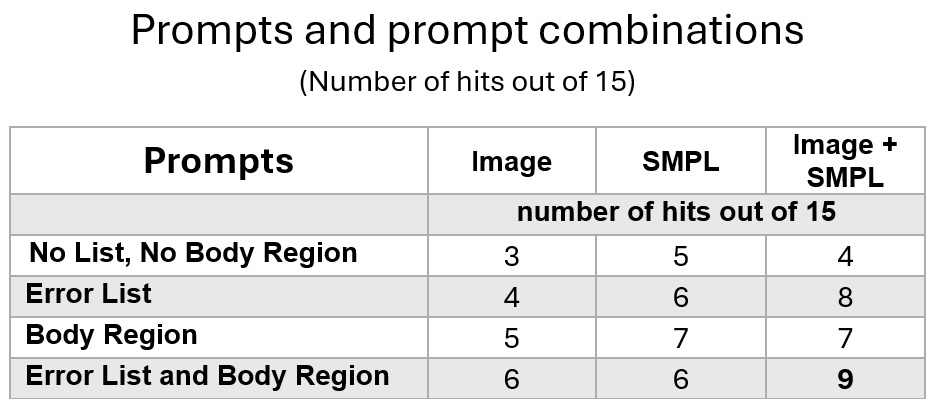}
    \end{tabular}
     \caption{Number of semantic matches with ground truths based on GPT-4 Vision for different prompt combinations (see text for details).}
    \label{tab:accuracies}
\end{table}

We compared GPT-4 Vision with open source tool VLMs. Various recent open-source models were included in these tests. Here, we show results on (i) Phi-3-vision-128k-instruct, (ii) InternLM-XComposer-2.5-OL, and (iii) Llava V1.6 Mistral 7b Hf (Table~\ref{fig:performance_cross_comp}).  The models were selected based on the results of various benchmarks \citep{zhao2024benchmarking} and included the GPT-4 Vision model, too.

We used LlaMa3 embeddings for the VLM model outputs with optimal prompts and computed cosine similarities with the ground truth vector. Results (2nd column of Table~\ref{fig:performance_cross_comp}) show GPT-4 Vision leading by a significant margin. We also tested whether the output matched the ground truth meaning. LlaVa1.6 is strict, scoring GPT highest, while the more permissive Phi3 Vision yields different results. InternLM stands out by giving non-zero scores for all VLMs except itself, with Phi-3-vision-128k-instruct scoring highest for InternLM. Evaluations of truth values are weak, except for GPT-4 Vision, but improvements are expected by new models that keep appearing.

\begin{table}
\begin{tabular}[h]{c}
    \includegraphics[width=0.94\linewidth]{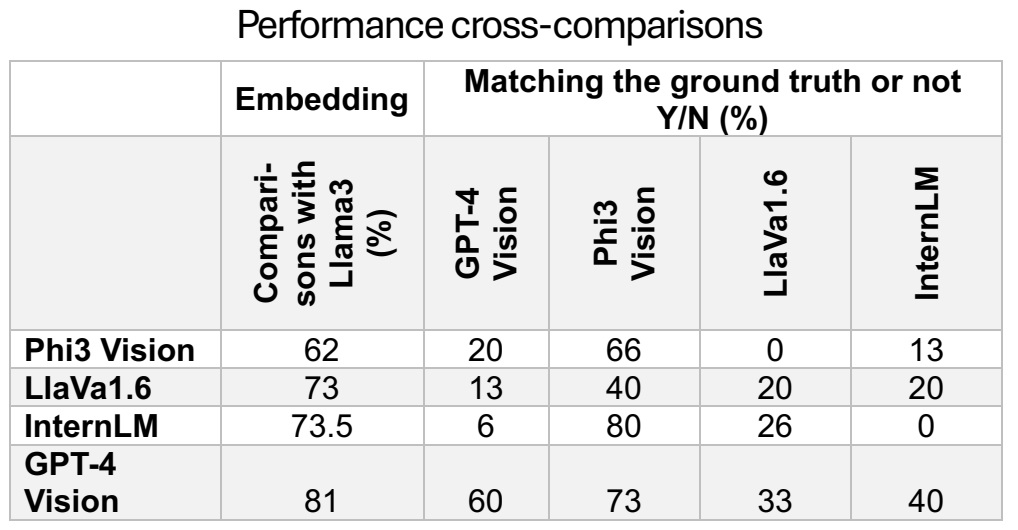}
\end{tabular}
\caption{Left column: methods. The second column shows cosine similarities between LLaMA3 embeddings of the ground truth and respective methods. Other columns show VLM results as percentages indicating matches or differences in meaning with the ground truth.}
\label{fig:performance_cross_comp}
\end{table}

\section{Discussion}\label{sec5}

The presented AIRS framework, a comprehensive solution for home physical rehabilitation. By integrating cutting-edge technologies such as real-time 3D reconstruction, VLMs, and advanced user modeling and path-planning methods, the article addresses significant challenges to overcome bottlenecks of rapidly growing needs in telerehabilitation.

\subsection{Strengths and Contributions}
\begin{description}
    \item[Innovative use of technology:] The AIRS framework effectively combines multiple technologies, such as VLMs for feedback, RT-3DR for spatial modeling, and smartphone-based guidance, providing a robust foundation for rehabilitation applications.
    \item[Personalization and Optimization:] By creating detailed 3D models of their living spaces and themselves, this framework guides users to set up their exercises, determine the optimal position of the camera and their own pose.
    \item[Focus on accessibility:] The framework is particularly inclusive, offering solutions for people with sensory impairments (e.g. blind patients or patients with hearing impairment) by integrating adaptable and accessible features.
    \item[Data-driven evaluation:] The study uses a robust dataset of 263 videos covering various exercises and errors, providing a strong basis for the evaluation of the effectiveness of the framework.
    \item[Privacy by Design:] The incorporation of patient avatars and anonymized data is in accordance with the European Union AI Act, ensuring privacy and ethical compliance.
    \item[Error detection:] The framework achieves an impressive 89\% detection rate for exercise errors using single-camera setups, with potential for improvement through RGBD recordings.
    \item[Feedback:] AIRS provides visual feedback by synchronizing the clinic's perfect exercise with the user's actual performance. Our results show that GPT-4 Vision helps therapists identify errors and offers correction suggestions. With the rapid advancement of VLMs, we expect a significant increase in correct suggestions, currently around 60\%. While corrective instructions are available, engaging in dialogue improves comprehension and adherence.
\end{description}

\subsection{Limitations and Challenges}

\begin{description}
    \item[Undetected Errors:] While the detection rate is high, 29 of 263 cases were challenging due to subtle errors, such as minor balance problems. This highlights the limitations of single-camera setups and the need for advanced recording methods. Two-camera recording seems sophisticated for home applications, but smartphones with RGBD cameras could provide a viable solution. Warnings before the execution of exercises that have hard-to-detect errors could also help. 
    \item[VLM Feedback Quality:] While VLMs like GPT-4 Vision offer useful feedback, their effectiveness is inconsistent, requiring therapist intervention and highlighting the need for further refinement.
    \item[Real-Time Constraints:] Real-time 3D modeling and localization require significant computational resources. Backing servers and streaming can solve the problem for low-power devices. 
    \item[Generalization:] The focus of the study on TKR exercises may raise questions about the applicability of the framework to other rehabilitation scenarios. While the motion patterns in these exercises are general, specific aspects, such as whether a particular muscle is being stretched (e.g., the thigh muscle in TKR exercises), are currently beyond the framework's capabilities.
\end{description}

Future directions include (a) extending data collection beyond TKR, (b) testing the framework with patients in real-world settings to assess usability, accessibility, and long-term outcomes, and (c) broadening the application domain to general exercises, which are typically faster, involve a greater variety of poses, and may also require high-quality RGBD detection.

\section{Conclusion and Outlook}\label{sec6}

\section{Contributions}
GB: conceptualization, investigation, methodology, 3D reconstruction methods, human pose estimation methods, data collection, supervision, implementation, writing, figures,
ZC: visual language model evaluation, prompt engineering, data processing,implementation and writing,
ÁF: supervision, mobile application testing,
KF: project administration, writing, reviewing, and editing
ZG: Rehabilitation dataset creation, exercise descriptions,
AL: conceptualization, methodology, project administration, supervision, and writing, reviewing, and editing,
JS: localization, path planning, navigation, implementation
All authors have read and agreed to the published version of the manuscript.

\section{Acknowledgment}
This research was supported by the European Union project RRF-2.3.1-21-2022-00004 within the framework of the Artificial Intelligence National Laboratory.
The authors also thank Robert Bosch Ltd., Budapest, Hungary for their generous support to the Department of Artificial Intelligence.
This project has also received funding from the European Union’s Horizon 2020 research and innovation programme under grant agreement No 952026. The support is gratefully acknowledged.
We would like to thank the hard work of Gábor Ambrus and Dávid Tokár on DeepRehab mobile application development as part of their BSc theses, under the supervision of Áron Fóthi and Gábor Baranyi, respectively.

\bibliography{sn-bibliography}% common bib file

\end{document}